# Demise of Faint Satellites around Isolated Early-type Galaxies

Changbom Park[1], Ho Seong Hwang[2,*], Hyunbae Park[3], and Jong Chul Lee[3]

[1]School of Physics, Korea Institute for Advanced Study, Hoegiro 85, Seoul 02455, Korea

[2]Quantum Universe Center, Korea Institute for Advanced Study, Hoegiro 85, Seoul 02455, Korea; hhwang@kias.re.kr, *corresponding author

[3]Korea Astronomy and Space Science Institute, 776 Daedeokdae-ro, Yuseong-gu, Daejeon 34055, Korea

**The hierarchical galaxy formation scenario in the Cold Dark Matter cosmogony with a non-vanishing cosmological constant and geometrically flat space has been very successful in explaining the large-scale distribution of galaxies. However, there have been claims that the scenario predicts too many satellite galaxies associated with massive galaxies compared to observations, called the missing satellite galaxy problem[1-3]. Isolated groups of galaxies hosted by passively evolving massive early-type galaxies are ideal laboratories for finding the missing physics in the current theory[4-11]. Here we report from a deep spectroscopic survey of such satellite systems that isolated massive early-type galaxies with no recent star formation through wet mergers or accretion have almost no satellite galaxies fainter than the r-band absolute magnitude of about $M_r = -14$. If only early-type satellites are used, the cutoff is at somewhat brighter magnitude of about $M_r = -15$. Such a cutoff has not been found in other nearby satellite galaxy systems hosted by late-type galaxies or those with merger features. Various physical properties of satellites depend strongly on the host-centric distance. Our observation indicates that the satellite galaxy luminosity function is largely determined by the interaction of satellites with the environment provided by their host, which sheds light on the missing satellite galaxy problem.**

To resolve the tension between the current galaxy formation theory and the observations of satellite galaxies it is necessary to find the missing physics by studying many dynamically well-defined galactic satellite systems. We select the isolated early-type host galaxies in the main survey region of the Sloan Digital Sky Survey (SDSS)[12]. Among the early-type galaxies (hereafter ETGs) having absolute magnitude of $M_r = -20.0 \sim -22.0$, and redshift of $z<0.011$ we selected those having no neighbor within $\pm 900$ km s$^{-1}$ along the line-of-sight and within the virial radius across the line-of-sight. Here neighbors are defined as galaxies having absolute magnitude not fainter by more than one magnitude and having redshifts within $\pm 900$ km s$^{-1}$ with respect to the target galaxy. According to these criteria seven isolated ETGs listed in Table 1 are selected. A flat cold dark matter universe having a non-vanishing cosmological constant ($\Lambda$CDM) with $\Omega_m = 0.3$ and $H_0 = 70$ km s$^{-1}$ Mpc$^{-1}$ is adopted. The virial radii are calculated using the method of refs. 13 and 14. The host ETGs are not only isolated but also have no strong wet-merger features. Only NGC 4125 shows some post merger features, but does not have cold gas or recent wet-merger feature. Four of them have active galactic nuclei.

We identify the satellite galaxies associated with these ETGs using uniform and highly complete redshift data taken from the literature and our deep spectroscopic survey. Since NGC 4125 and NGC 5363 are located relatively at closer distances, we use the redshifts of surrounding galaxies obtained from the literature for these systems. For the remaining five systems we perform a spectroscopic survey of galaxies located within half a degree from each host galaxy and having the Galactic extinction-corrected SDSS data release 12 (DR12) magnitude of $m_{r,Petro,0}$< 20.5. No colour selection is applied. The SDSS DR12 spectroscopic data archive[15] provides nearly a complete set of galaxy redshifts down to $m_r$=17.77, and our observation extended the survey depth by 1.7-2.6 magnitude (see Table 1). The observation has been made with Hectospec on MMT from 2014 through 2016. The Hectospec is a 300 optical fiber fed spectrograph with a 1 degree field of view[16]. A total of 4,312 redshifts are measured in the fields of five systems. We combine these data with those from the SDSS DR12 and from the NASA/IPAC Extragalactic Database (NED). Satellites are identified by using their proximity to their hosts along and across the line of sight. The details of satellite identification are presented in the Methods section. Absolute magnitudes of the identified 49 satellites are plotted in the top panel of Figure 1 for five hosts for which the survey limits in absolute magnitude (vertical bars) are fainter than $M_r$=-13.

We calculate the satellite galaxy luminosity function (hereafter LF) taking into account that the survey limit is different for different hosts and the survey completeness varies as apparent magnitude. The LF $\phi(M_r)$ is estimated from the number of satellites with the absolute magnitude in a bin ($M_r$-0.5$\Delta M_r$, $M_r$+0.5$\Delta M_r$) as in

$$\phi(M_r)\Delta M_r = \sum_{i=1}^{N_{sys}} w_i(M_r)^{-1} N_{sat,i}(M_r,\Delta M_r) / \sum_{i=1}^{N_{sys}} V_{survey,i}, \qquad (1)$$

where $N_{sys}$ is the number of the satellite systems within the survey limit in a $M_r$ bin, and $w_i(M_r)$ is the survey completeness at $M_r$ for the i-th system. $V_{survey,i}$ is the survey volume for the i-th hosts when the survey limit is taken into account. The result is shown in the middle panel of Figure 1. The surface brightness incompleteness in the SDSS photometric catalog has been corrected (open and filled circles are the LFs before and after the correction). The satellite galaxy LF rises as $M_r$ increases to about -15, but clearly drops steeply at magnitudes fainter than $M_r$=-14.

In the bottom panel morphologically early- and late-type satellites are plotted as red circles and blue stars, respectively, as a function of host-centric distance. It indicates that the cutoff is at a brighter magnitude of about -15 for early-type satellites. It should be noted that the late-type satellites should show such a cutoff at a brighter magnitude if the cutoff has appeared due to incompleteness of our observational sample for faint low surface brightness galaxies as their surface brightness is in general lower than that of early-type satellites. The cutoffs seem to be insensitive to the host-centric distance. The figure also shows that the satellite galaxy LF depends on the distance from the host galaxy. As the distance increases, there are relatively fewer bright satellites.

We have made a double check to confirm that the existence of the cutoff is not affected by observational biases by comparing the LFs of the satellites and the field SDSS galaxies. The field galaxy LF is measured from the deep blind redshift surveys of SHELS F1/F2 and GAMA G15 (refs. 17 – 19). We select the galaxies in these surveys from the SDSS photometric catalog, and apply the same surface brightness incompleteness correction to the resulting LF. We also apply the spectroscopic incompleteness correction. Only the galaxies at 0.002<z<0.008 are used as in our observational sample. We adopt the apparent magnitude limits of $m_r$=20.5 for F1 and F2, and $m_r$=19.4 for GAMA G15. The corresponding difference in survey depths has been taken into account in LF calculation at each absolute magnitude bin. Results are the black dots in Figure 2. There is no cutoff in the field galaxy LF down to $M_r$=-13 unlike the LF of the satellite galaxies even though galaxies are drawn from the same photometric catalog.

We have also estimated the statistical significance of the cutoff. We generated 1000 samples of random galaxies with $-22.0<M_r<-12.5$ located at $0.002<z<0.008$. We sample galaxies randomly according to the field galaxy LF given in Figure 2, and apply the observational selection effects. Each sample is constrained to have the same number of galaxies as in our satellite galaxy. It turns out that only 0.5% of those samples have the ratio $\phi(-14.0<M_r<-12.5)/\phi(-15.5<M_r<-14.0)$ equal to or smaller than the observed value of 0.24. This tells that the satellite galaxy LF is different from the field galaxy LF and the observed cutoff is statistically significant.

We further examine in the Methods section whether or not the satellites in the combined system of Centaurus A and the Milky Way would show an artificial cutoff near $M_r \approx -14$ if the system were observed at the distance of NGC 3665, the farthest system used for constructing the satellite galaxy LF in our study. There is no such effect. This demonstrates that the cutoff found for our systems is not produced by an observational bias such as incomplete sampling of faint low surface brightness galaxies.

Figure 3 shows how the stellar mass, g-r colour, and Sérsic index of satellite galaxies in the seven systems in Table 1 depend on the distance from host galaxies, $R_{sat}$. Relatively more massive satellites are concentrated to the central region of the systems regardless of satellite morphology. The phenomenon is likely due to the mass dependence of the merger time scale as more massive satellites have shorter merger-time scale[20]. The middle panel shows that the dispersion in colour increases as separation decreases, and a red sequence of both early- and late-type satellites becomes redder toward smaller $R_{sat}$. This colour dependence of satellite galaxies on host-centric distance is clear evidence for the impact of host galaxies on the physical state of their satellites. The bottom panel shows that most satellites at $R_{sat} \geq 0.4 r_{vir,h}$ are disky galaxies. But the surface brightness of some satellites appear centrally concentrated with large Sérsic indices when $R_{sat} \sim 0.2\ r_{vir,h}$. About half of them have starbursts at the center that may have been triggered by interactions with host or other satellites. At $R_{sat}<0.1 r_{vir,h}$ some satellite galaxies appear disturbed by the tidal force of their host making Sérsic index small. All these dependence of satellite galaxy properties on host-centric distance supports that various physical properties of satellite galaxies depend sensitively on the environment provided by their host.

On the other hand, it should be pointed out that early-type galaxies can have satellites fainter than the cutoff if they have recently experienced a wet merger and formed or accreted faint galaxies. An example is Centaurus A (Cen A), which is an ETG and satisfies our isolation criteria but has many satellites fainter than the cutoff found for our sample. The peculiar galaxy, Cen A, is very likely to have experienced a recent wet merger event evidenced by its complex merger features[21] and two planes of satellites[22]. Another example would be Sombrero galaxy, a peculiar galaxy that looks like a composite object of a massive elliptical galaxy and a thin-disk spiral galaxy. The galaxy's dusty disk containing many young bright stars may be evidence for a recent wet merger or accretion. Because of this complication Sombrero is again not an ideal laboratory to examine whether or not the hostile environment provided by the host removed faint satellites.

The resolution to the missing satellite galaxy problem might come from better understanding of baryonic physics, dark matter physics, and gravitational force, or might require modification of the background cosmology. But the fact that faint satellites below $M_r \approx -14$ are missing in the satellite galaxy system hosted by an ETG while there is no such a cutoff for the systems hosted by star-forming galaxies indicates that the environment provided by the massive early-type host galaxy is responsible for the cutoff. Therefore, the observed cutoff in the satellite galaxy LF of isolated bright ETGs provides evidence for the critical role of baryonic physics in the evolution of satellite galaxies and can be a clue to resolve the missing satellite galaxy problem.

**Acknowledgements**

We thank Korea Institute for Advanced Study for providing computing resources (KIAS Center for Advanced Computation Linux Cluster System) for this work.

**Author Contributions**

C.P. led the project and wrote most of the text.

H.S.H. made the spectroscopic survey and data analysis.

H.P. analyzed the simulation data for comparison with observation.

J.C.L. measured the stellar mass and Sérsic index of satellite galaxies.

**Author Information**

Reprints and permissions information is available at www.nature.com/reprints. The authors declare no competing financial interest. Correspondence and requests for materials should be addressed to H.S.H. (hhwang@kias.re.kr).

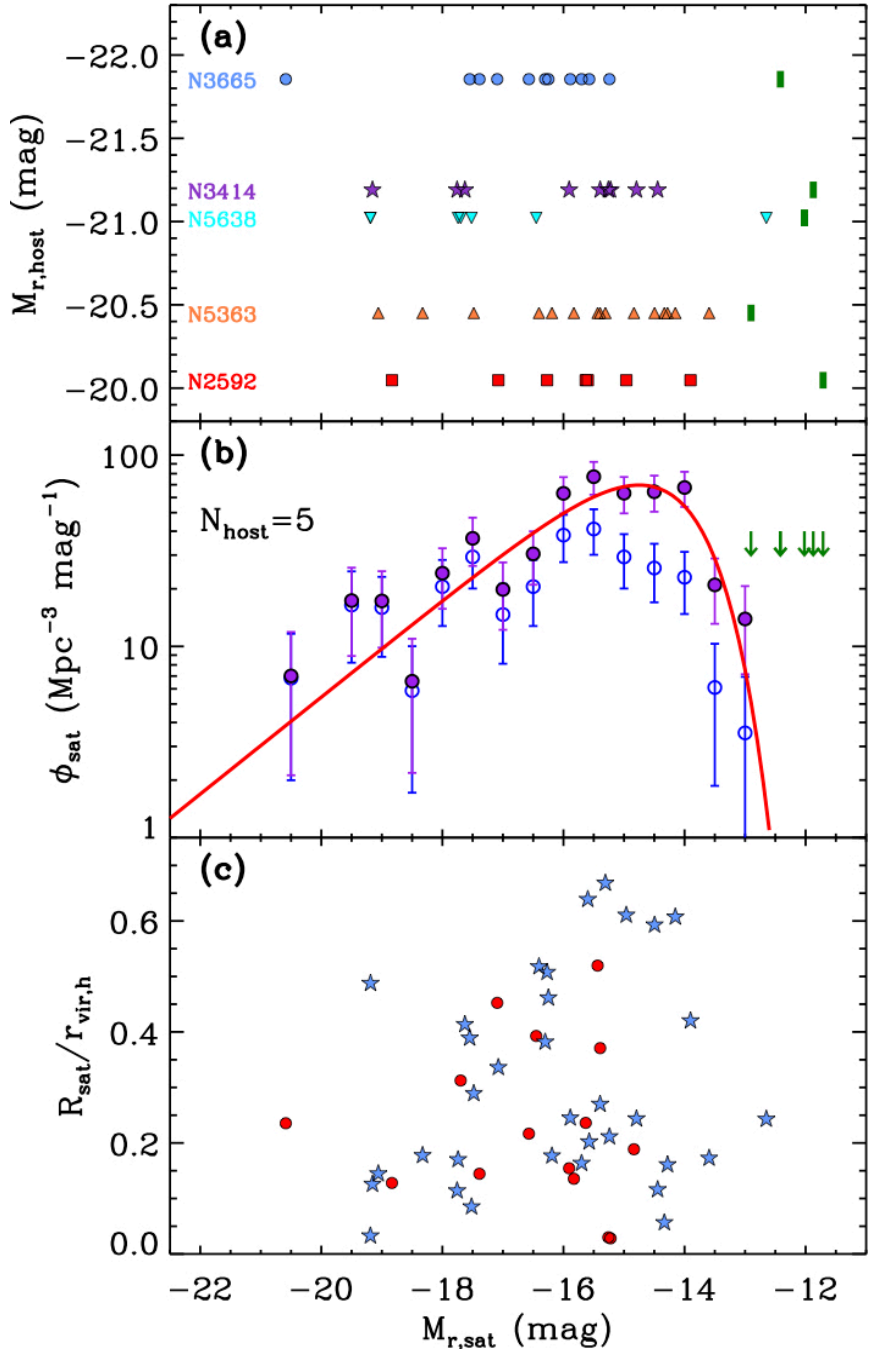

**Figure 1**. **Physical properties of satellite galaxies as a function of absolute magnitude. a,** Absolute magnitudes of host galaxies. The green vertical bars mark our survey limits in absolute magnitude for individual hosts. **b,** Luminosity function of the satellite galaxies in five systems (NGC 2592, NGC 3414, NGC 3665, NGC 5363, NGC 5638) showing a cutoff near $M_r = -14$. Open and filled circles are the functions before and after correction for the surface brightness incompleteness. The error bars correspond to the Poisson error. Vertical arrows mark the survey limits in the five fields. The solid line is a function $\phi(L) = \phi_0 (L/L_*)^\alpha \exp[-\beta(L/L_*)]$ best fit to the data. **c,** Host-centric distance of the satellites normalized by host virial radius. Red points are early-type satellites, and blue points are late-type satellites.

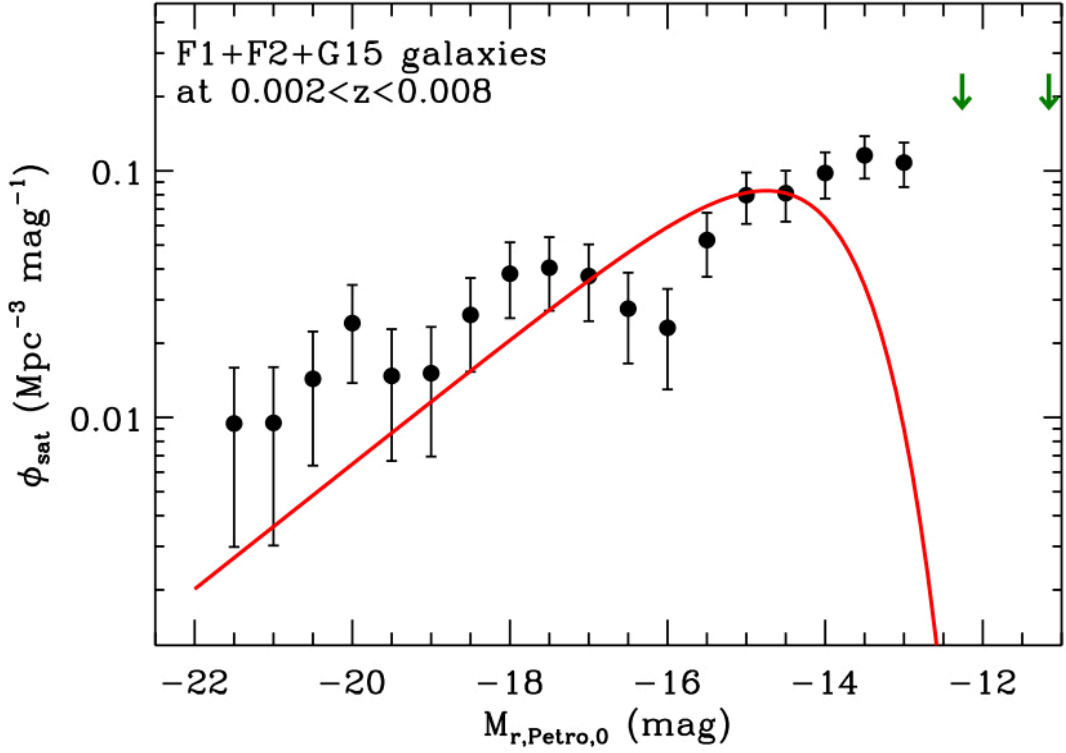

**Figure 2**. **The luminosity function of the field galaxies in the GAMA G15 survey[19] and the SHELS F1 and F2 surveys[17,18]**. The error bars correspond to the Poisson error. The left and right arrows mark their survey limits, respectively. The corrections for the surface brightness incompleteness in the SDSS photometric catalog and for the spectroscopic survey incompleteness have been made. The red curve is the best-fit satellite galaxy luminosity function shown in Figure 1, matched with the field galaxy LF at $M_r$=-14.5.

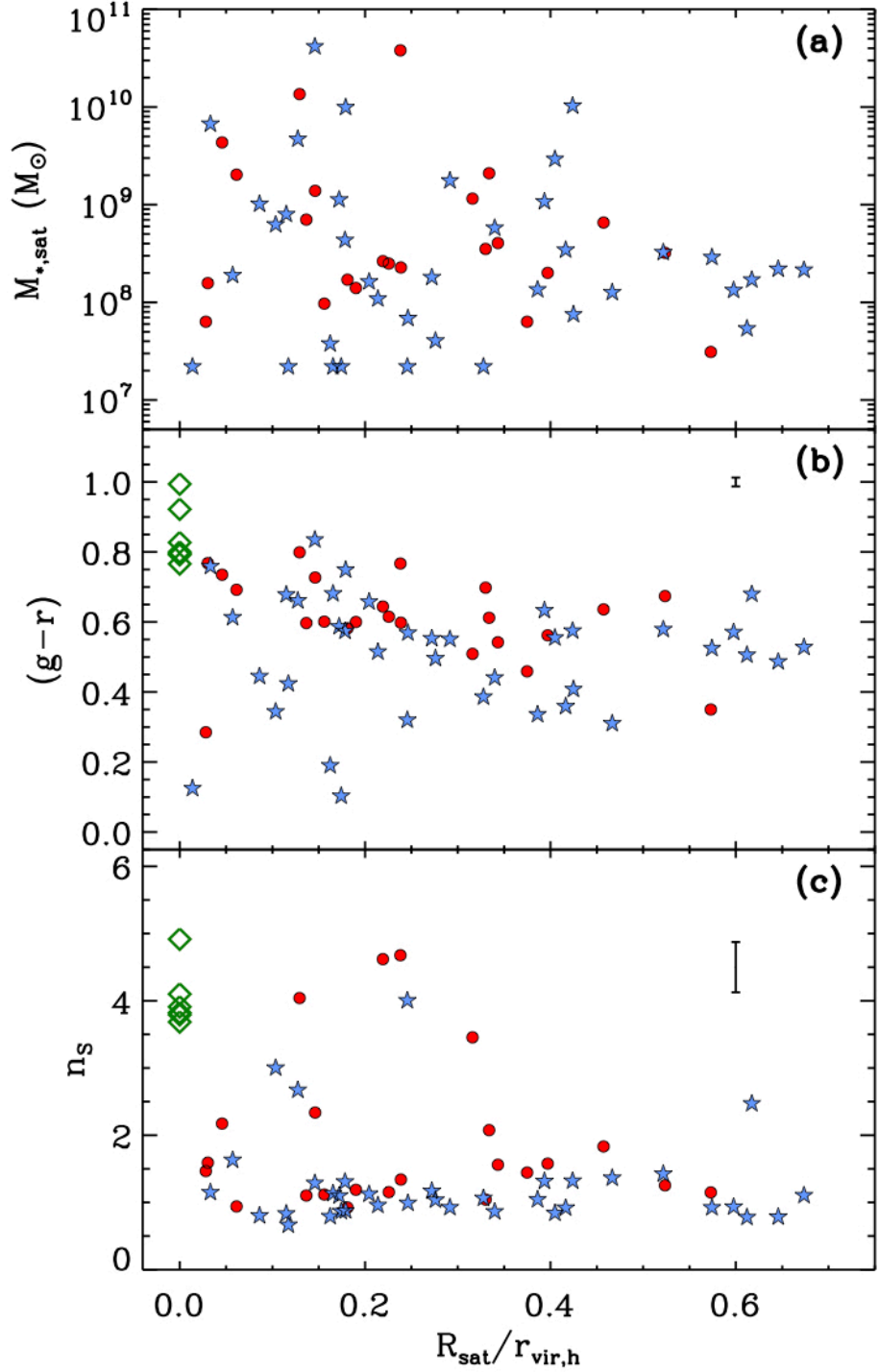

**Figure 3**. **Physical parameters of galaxies in the seven satellite galaxy systems of Table 1 as a function of host-satellite distance in units of host virial radius. a,** Stellar mass. **b,** g-r colour. **c,** Sérsic index. Red and blue symbols are early- and late-type satellites, respectively. Green diamonds represent host galaxies. Typical errors are represented with an error bar.

**Tables**

Table 1. Isolated early-type host galaxies selected in this work.

| NGC | RA[1] | DEC[1] | z | D[2] | $m_r$[3] | $M_r$ | $\log M_*$[4] | T | $r_{vir}$[5] | $m_{r,lim}$[6] | $M_{r,lim}$[7] | $N_{sat}$[8] |
|---|---|---|---|---|---|---|---|---|---|---|---|---|
| 2592 | 126.78354 | 25.97031 | 0.006825 | 26.3 | 12.205 | -20.05 | 10.60 | E2 | 346 | 20.38 | -11.71 | 7 |
| 3414 | 162.81754 | 27.97510 | 0.004903 | 25.7 | 10.918 | -21.19 | 10.87 | S0p | 493 | 20.17 | -11.87 | 9 |
| 3665 | 171.18196 | 38.76279 | 0.006835 | 33.1 | 10.786 | -21.85 | 11.22 | S0 | 604 | 20.17 | -12.42 | 11 |
| 3872 | 176.45437 | 13.76668 | 0.010627 | 44.7 | 11.614 | -21.77 | 11.30 | E5 | 588 | 19.47 | -13.77 | 10 |
| 4125 | 182.02453 | 65.17434 | 0.004523 | 24.0 | 10.258 | -21.69 | 11.15 | E | 573 | 17.77 | -14.11 | 6 |
| 5363 | 209.02998 | 5.25503 | 0.003799 | 13.6 | 10.283 | -20.45 | 11.22 | E/S0p | 393 | 17.77 | -12.90 | 15 |
| 5638 | 217.41823 | 3.23333 | 0.005591 | 26.5 | 11.178 | -21.02 | 10.91 | E1 | 468 | 20.09 | -12.02 | 7 |

1. RA and DEC in degrees
2. Luminosity distances in units of Mpc.
3. Petrosian apparent magnitude in the r-band.
4. Stellar mass in units of $M_\odot$.
5. Virial radius in kpc.
6. Limiting magnitude where the completeness of the spectroscopic survey falls to 50%.
7. Limiting absolute magnitude corresponding to the limiting apparent magnitude.
8. Number of satellites identified in this work.

# Methods

**Observation and data analysis:**

The redshift survey of galaxies was made with Multiple Mirror Telescope. We used the 270 line mm$^{-1}$ grating of Hectospec that provides a dispersion of 1.2 Å pixel$^{-1}$ and a resolution of about 6 Å (R =1000). The resulting spectral coverage is 3700-9100 Å. We observed 2-5 fields for each satellite system, which centered on the host galaxy (see Table 1). Thus the observational bias that can be introduced by fiber collisions is minimized. Typical on-source-exposure per field is 45-60 minutes. We reduced the Hectospec spectra with HSRED v2.0 that is an updated reduction pipeline originally developed by Richard Cool. We then used the RVSAO[23] to determine the redshifts by cross-correlating the spectra with templates. To avoid the bias introduced by using a small number of templates, we used 9 templates with various types that include the spectra of composite elliptical and spiral galaxies, composite absorption- and emission-line galaxies, an SDSS QSO and three M31 globular clusters. Among the 9 redshift estimates for a given object, we adopted the one with the highest cross-correlation signal provided by the RVSAO (i.e. r-value that is an indicator of cross-correlation reliability[24]) as its best estimate. We finally adopted the redshift of a galaxy only when r is larger than 4, consistent with the limit confirmed by visual inspection[17,18]. Repeat observations of 1651 separate absorption-line and 238 separate emission-line objects provide mean internal errors normalized by (1 + z) of 48 and 24 km s$^{-1}$, respectively[17]. The survey is summarized in Supplementary Table 1 where the final numbers of redshifts are given. Supplementary Figure 1 shows the objects with known redshifts within the survey region of each system.

**Distances to the satellite galaxy systems**

To obtain the absolute magnitude of the host and satellite galaxies it is necessary to know distances to the satellite systems. In Table 1 distances to NGC 2592, NGC 3414, NGC 4125, NGC 5364 (and NGC 5363), and NGC 5638 are the luminosity distances from The Extragalactic Distance Database (EDD)[25]. NGC 5363's distance is taken from that of NGC 5364, which is a satellite of NGC 5363. Distance to NGC 3665 is from ATLAS$^{3D}$ catalog[26]. Distance to NGC 3872 is derived from the mean redshift of the satellite-host system of z=0.01047 and using a flat ΛCDM cosmology with $\Omega_m$=0.3.

**Identification of satellites:**

Satellite galaxies are identified as follows. In Supplementary Figure 2 the satellite galaxy candidates are plotted in the plane of velocity difference $\Delta v$ from the mean group velocity and host-centric radius in units of the host virial radius $R/r_{vir,h}$. To reduce the potential effects of interlopers we put reasonable boundaries separating the satellites from interlopers using the following formula

$$\Delta v < v_{esc} = v_0 10^{-0.4(M_{r,h}-M_0)/3}(R/r_{vir,h})^{-1/2},$$

where $M_{r,h}$ is the absolute magnitude of the host. We adopt $v_0$=250 km s$^{-1}$ for $M_0$=-21.0. Our results are hardly affected by different choices of $v_0$. In addition to the constraint on velocity difference we also adopt the constraint R< 0.7 $r_{vir,h}$ to reduce interlopers further and to see the effects of host on satellites more clearly.

**Correction for surface brightness incompleteness**

The SDSS photometric catalog, which was used for our spectroscopic survey target list, is known to be increasingly incomplete for faint objects. This is because galaxies with low luminosity tend to have low surface brightness and are likely to be missing in the SDSS photometric catalog[27]. We used the correction factor for the surface brightness incompleteness given in Fig. 6 of ref. 27. The filled circles in the middle panel of Figure 1 are the LF after the correction. The cutoff in the LF remains to exist at around the same magnitude even after the correction.

**Luminosity function of the satellite galaxies of Centaurus A and the Milky Way**

We have conducted the following experiment using the known satellite galaxies of Cen A and the Milky Way (MW) to check whether or not the cutoff is caused by some observational biases. We first compile the photometric data of the satellite galaxies of Cen A from refs. 21 and 28. Their V-band magnitudes are converted to the SDSS r-band magnitudes by using the relation $m_r \sim m_V + 0.215$, which is derived from the galaxies common in ref. 28 and SDSS. Similarly, the data for the MW satellites is obtained from the updated version of nearby galaxy catalog of McConnachie[29]. The absolute r-band magnitudes of all the satellites are calculated. The satellite galaxy LFs are plotted in Supplementary Figure 3 for the MW (triangles with dashed line), Cen A (squares with dot-dashed line), and the combined system (red circles with error bars). The plot shows that the satellite galaxy LF of Cen A plus the MW system does not show a cutoff down to $M_r = -12$. Now the question is whether or not one finds a spurious cutoff if this system is observed at the distance of our early-type hosts due to the observational selection effects.

To answer the question we made the following experiment examining the impact of the surface brightness incompleteness effect on the satellite galaxy LF. We first compile the surface brightness data of the satellite galaxies of Cen A from refs. 21 and 28, and of MW from ref. 29. We convert them into the SDSS magnitude system as above. We then construct a subsample of galaxies by randomly selecting the satellites following the surface brightness incompleteness curve provided by ref. 27 for the SDSS photometric catalog (see their Fig. 3). We then apply the selection effect of spectroscopic survey incompleteness as a function of apparent magnitude. We assume that the satellites are at the distance of NGC 3665 or D=33.1 Mpc, and calculate their new apparent r-band magnitudes. NGC 3665 is the farthest system used for constructing the satellite galaxy LF in our study. The effective completeness of the mock sample at $M_r = -14$, -13, and -12 at D=33.1 Mpc is 0.321, 0.228, and 0.025, respectively. The completeness at magnitudes brighter than and equal to $M_r = -13$ is still not too low and the LF can be restored, but at $M_r = -12$ the completeness begins to be so low that our correction for the incompleteness can fail unless the number of galaxies is very large. We finally calculate the LF after making corrections for incompleteness due to the surface brightness and the spectroscopy. Because the number of satellites in a magnitude bin is usually small, we do this experiment 1000 times so that the overall sampling probability follows well the incompleteness curves. The blue solid line is the average LF determined from 1000 calculations.

The figure shows that the LF of the satellites in the Cen A plus MW system can be restored quite well down to $M_r = -13$ for SDSS-like surveys even if they were located as far as 33.1 Mpc. This experiment demonstrates that the cutoff we discovered around $M_r = -14$ for the early-type host systems is genuine and not produced by observational biases.

**Data Availability Statement**

The data that support the plots within this paper and other findings of this study will be published are available from the corresponding author upon reasonable request.

**Supplementary information** is available for this paper at doi:

# Demise of Faint Satellites around Isolated Early-type Galaxies

# (Supplementary information)

Changbom Park, Ho Seong Hwang, Hyunbae Park, & Jong Chul Lee

**This document contains one supplementary table and four supplementary figures.**

**Tables:**

**Supplementary Table 1**. Spectroscopic survey of galaxies for five systems

| Hosts | SDSS # | Trimester | Exposures(min) | N(z)[1] |
|---|---|---|---|---|
| NGC 2592 | 1142165 | 2016a | 45/60/60/60 | 860 |
| NGC 3414 | 2204882 | 2016a | 45/60/60/60 | 897 |
| NGC 3665 | 1317351 | 2014a | 60/60 | 447 |
| | | 2015b | 60/60 | 508 |
| NGC 3872 | 1122257 | 2016a | 45/60 | 495 |
| NGC 5638 | 485766 | 2015b | 45/60/60/60 | 961 |
| | | 2016a | 60 | 144 |

1. The number of galaxy redshifts obtained newly.

**Figures:**

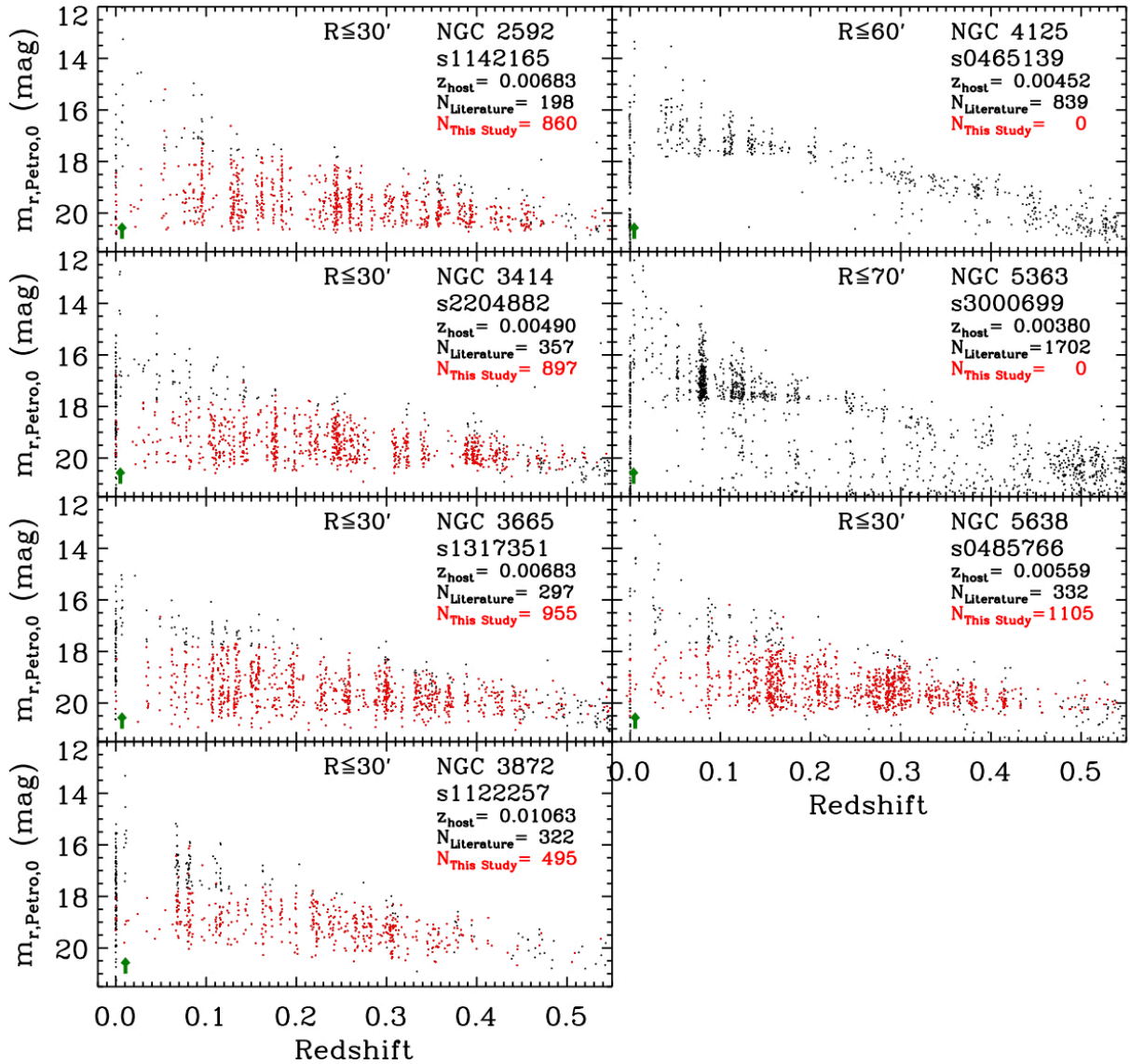

**Supplementary Figure 1**. **The galaxies with known redshifts within the survey region of each satellite system.** The y-axis is the r-band apparent Petrosian magnitude. A typical error in magnitude is smaller than the symbol size. Black points are the objects with redshifts in the literature and red points are those with newly observed redshifts. The satellite systems are marked as a green arrow in each panel. The left-most groups of points are stars in the Milky Way.

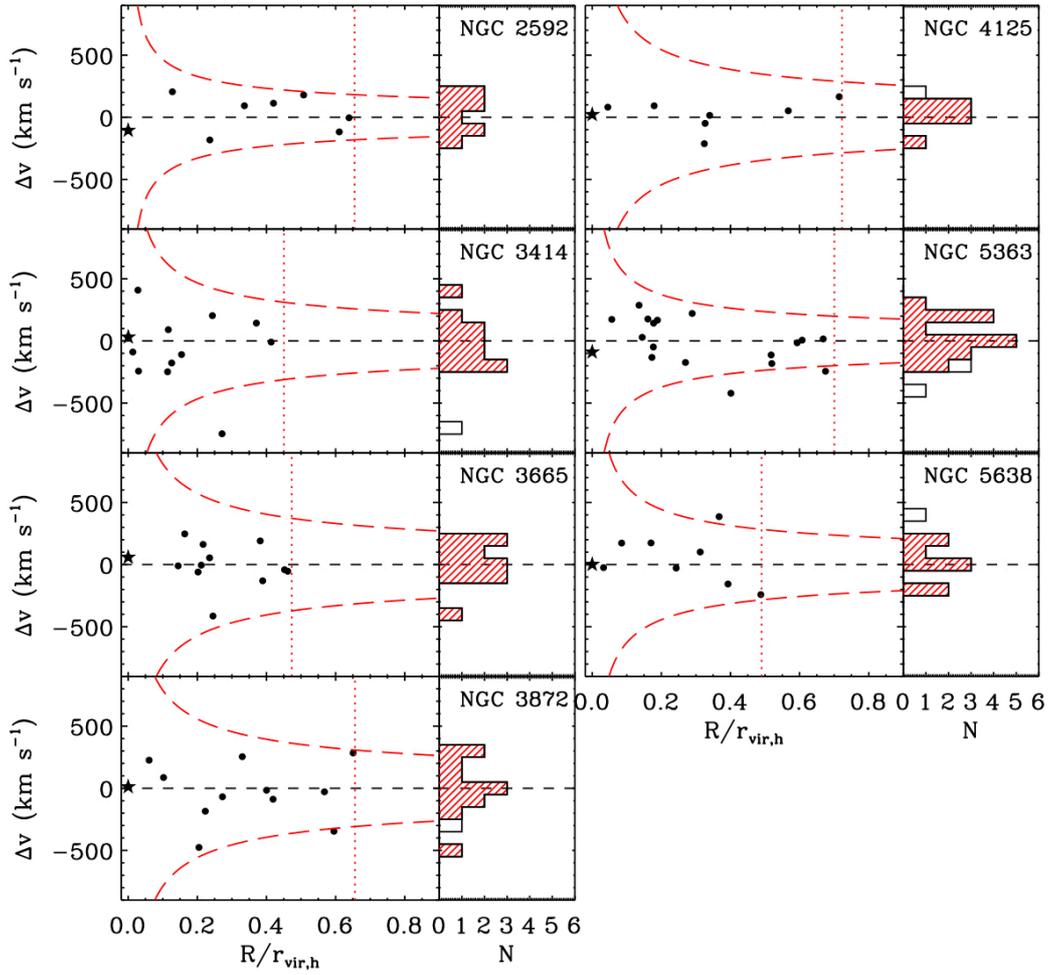

**Supplementary Figure 2**. **The velocity difference with respect to the mean group velocity of the satellite system members as a function of host-centric distance and its histogram.** The stars are the host galaxies, and points are satellite candidates. A typical error in velocity difference is about 60 km s$^{-1}$. The red long-dashed lines are our choice of the velocity difference constraint corresponding to $v_0$=250 km s$^{-1}$ for $M_0$=-21.0. The radii corresponding to 30' are marked by vertical dotted lines except for the more nearby galaxies, NGC 4125 and NGC 5363 for which the dotted lines indicate the angular separation of 60' and 70', respectively.

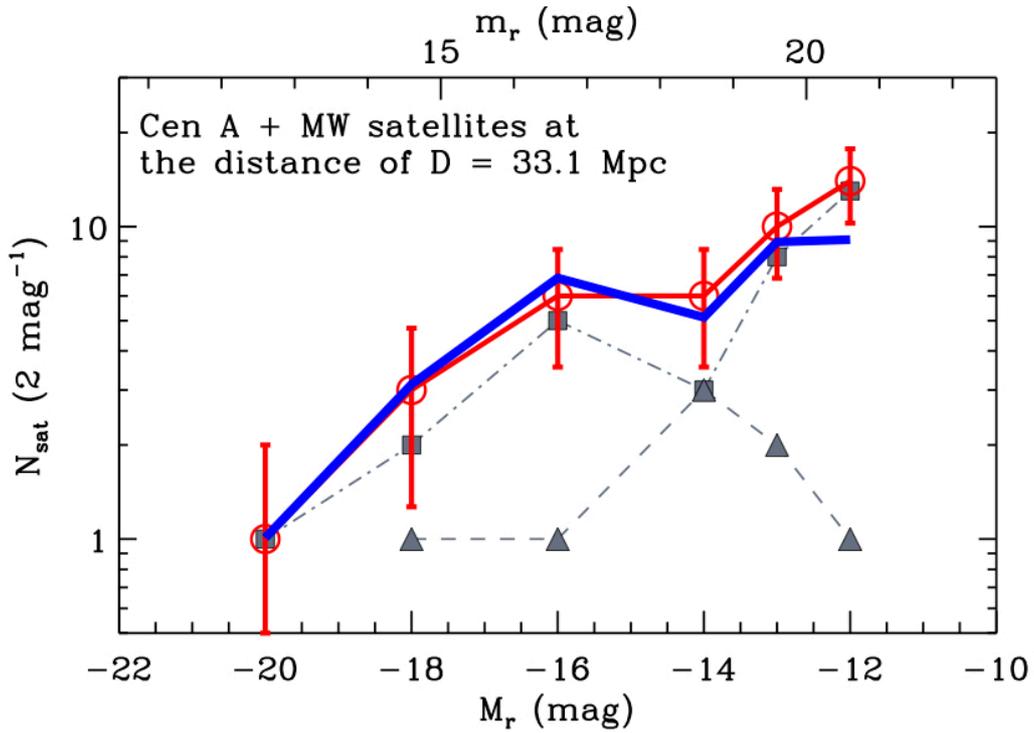

**Supplementary Figure 3. Luminosity function of the satellite galaxies of Centaurus A (squares with dot-dashed line) and the Milky Way (triangles with dashed line) down to $M_r$ =-12.** The red line with error bars is the combined LF and the error bars correspond to the Poisson error. The blue line is the average LF from 1000 experiments by applying the observational selection effects and making corrections to the effects. We assume that the satellites are at the distance of 33.1 Mpc  The reconstruction starts to fail at $M_r$ =-12.

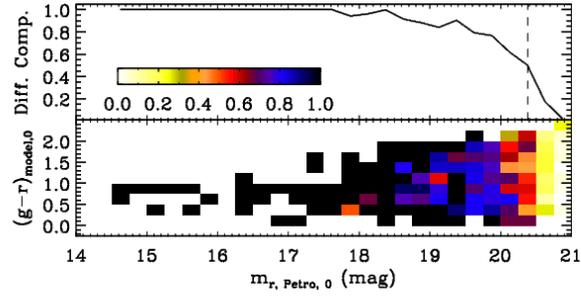
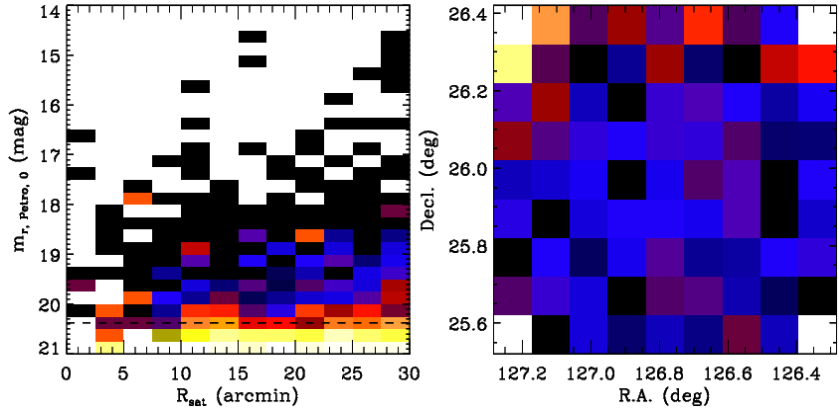
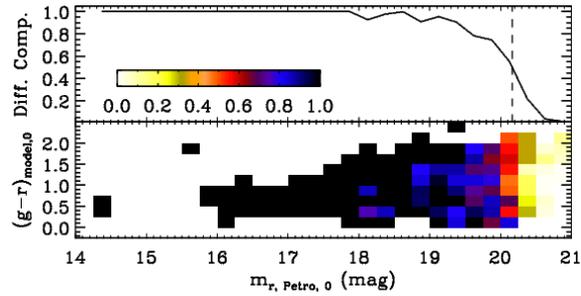
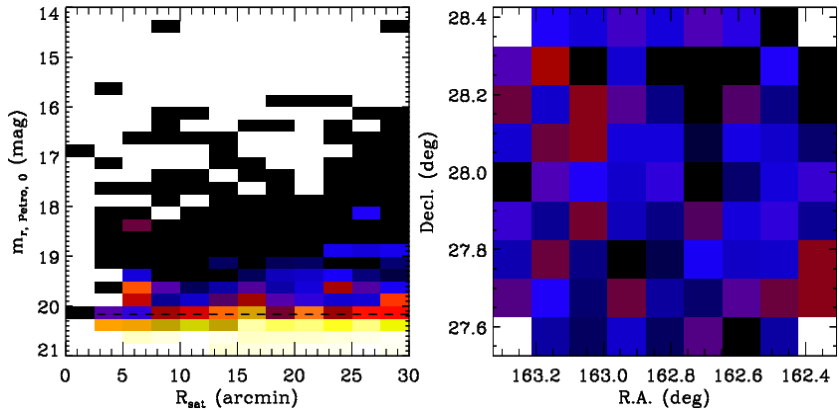

**Supplementary Figure 4**. **The completeness of our redshift survey.** Spectroscopic completeness as a function of apparent magnitude, colour (top panel), host-centric radius (bottom left) and on the sky (bottom right). In the bottom right panel, the colour represents the cumulative completeness down to each survey limiting magnitude. The host is located at the center of the panel.

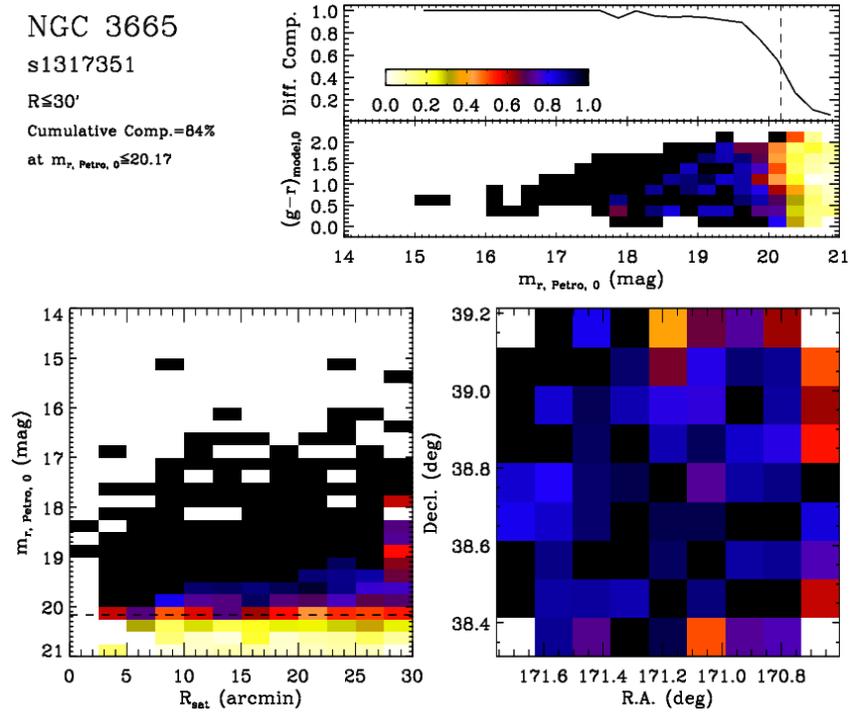
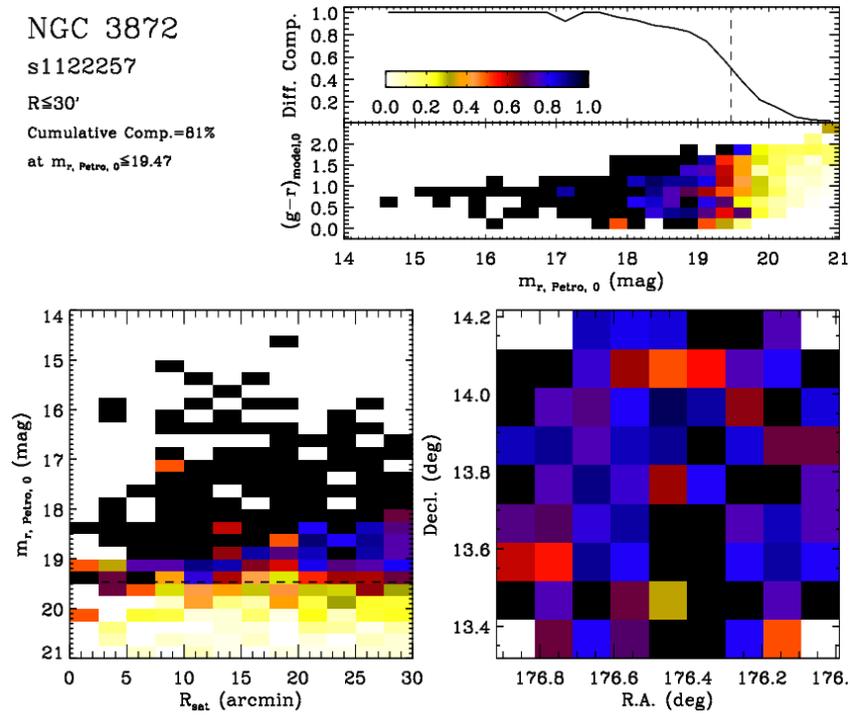

**Supplementary Figure 4** (cont.).

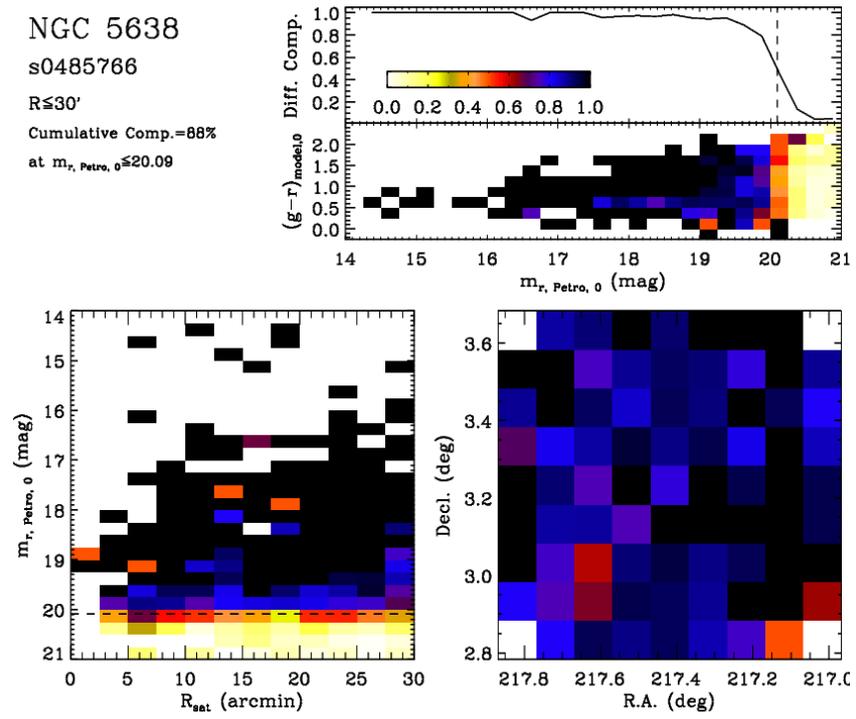

Supplementary Figure 4 (cont.).

**Host galaxies:**

The completeness of our redshift survey in apparent magnitude, colour, host-centric radius, and on the sky is shown in Supplementary Figure 4. For each satellite system, the top panel shows the differential spectroscopic completeness as a function of r-band magnitude at R≤30'. The spectroscopic completeness is the ratio of the number of galaxies with spectroscopic redshifts to the number of galaxies in the SDSS photometric catalog. The vertical dashed line indicates the limiting magnitude ($m_{r,lim}$) where the completeness falls to 50%. To examine the color dependence of the spectroscopic completeness, we also show the two-dimensional distribution of the completeness as a function of g-r colour and r-band magnitude in the following panel. The two bottom panels show the two-dimensional distributions of the spectroscopic completeness as a function of r-band magnitude and of projected host-centric distance (left panel) and as a function of R.A. and DEC. at $m_{r,Petro,0} \leq m_{r,lim}$ (right panel). The horizontal line in the bottom left panel indicates the limiting magnitude.

All of our host galaxies are listed as groups in the galaxy group catalog of ref. 11 in the main text. Additional information on the hosts of our seven galactic satellite systems is as follows.

NGC 2592 (or UGC 4411) is a typical elliptical galaxy with morphological type of E2. Its rotational velocity is 148 km s$^{-1}$ (ref. 1). Its nearest neighbor galaxy is a 0.62 magnitude fainter spiral galaxy (NGC 2068) separated by 1.6 Mpc across the line of sight.

NGC 3414 (or Arp 162, UGC 5959) is known as a peculiar galaxy with morphological type of S0p. It is a slow rotator with $V_{max}$=34 km s$^{-1}$ (refs. 1 and 2), and has a LINER-type AGN[3]. The mass of the supermassive black hole is estimated to be 2.5×10$^8$ M$_\odot$ (ref. 4). Its nearest neighbor is NGC 3504 which is a late-type galaxy with almost the same magnitude and is 0.94 Mpc away across the line of sight.

NGC 3665 is an E2 galaxy showing a double radio structure perpendicular to the inner thin dust lane[5]. It is rotating with $V_{max}$=149 km s$^{-1}$. The estimated mass of the central supermassive black hole is about 6×10$^8$ M$_\odot$ (ref. 6). Its nearest neighbor is NGC 3788, a spiral galaxy at the separation of 3.8 Mpc across the line of sight.

NGC 3872 has an E5 galaxy at z=0.0104 and is the farthest in our sample. NGC 3800, a 0.7 magnitude fainter spiral galaxy, is its nearest neighbor galaxy separated by 1.6 Mpc across the line of sight.

NGC 4125 (UGC 7118) is an E type galaxy. It has a LINER AGN producing hard X-ray[3]. It shows some post-merger features. Its nearest neighbor galaxy is NGC 4036 located at 1.14 Mpc across the line of sight.

NGC 5363 is a S0p or E-type galaxy showing an inner dust lane along the minor axis and an outer dust lane along the major axis[7,8]. It has a LINER AGN-type source producing X-ray, radio and IR emission[3,9-11]. It has the nearest neighbor NGC 5248 at 1.67 Mpc across the line of sight.

NGC 5638 is a clean E1 galaxy showing rotational motion with $V_{max}$=82 km s$^{-1}$. Its nearest neighbor galaxy is NGC 5576, which is an early-type galaxy brighter than NGC 5638 by 0.25 and is 1.0 Mpc away across the line of sight.